\newcommand{\be}{\begin{equation}}\newcommand{\ee}{\end{equation}}
\newcommand{\bea}{\begin{eqnarray}}\newcommand{\eea}{\end{eqnarray}}
\newcommand{\nn}{\nonumber\\}\newcommand{\p}[1]{(\ref{#1})}

\documentstyle[12pt]{article}

\begin{document}
\begin{titlepage}
\begin{flushright}
LPTHE 01-25\\
hep-th/0105210 \\
May, 2001
\end{flushright}
\begin{center}
{\Large \bf{Superbranes and Super Born-Infeld Theories as Nonlinear
Realizations}} \vspace{1.5cm} \\ E. Ivanov${\;}^{a,b}$ 
                    \vspace{1cm} \\
\centerline{${}^{a}$\it Bogoliubov  Laboratory of Theoretical Physics, JINR,}
\centerline{\it 141 980 Dubna, Moscow Region, Russia} \vspace{0.3cm} 
\centerline{${}^{b}$\it
Laboratoire de Physique Th\'eorique et des Hautes Energies,}
\centerline{\it Unit\'e associ\'ee au CNRS UMR 7589,~Universit\'e Paris 7}
\centerline{\it 2 Place Jussieu, 75251 Paris Cedex 05, France}\vspace{0.5cm}
\end{center}
\begin{abstract}
\noindent{We outline, on a few instructive examples, the characteristic
features of the approach to superbranes and super Born-Infeld theories based 
on the concept of partial spontaneous breaking of global supersymmetry (PBGS). 
The examples include the  $N=1, D=4$ supermembrane and the
``space-filling'' D2- and D3-branes. Besides giving a short account of the
available results for these systems, we present some new developments. For the
supermembrane we prove the equivalence of the equation of motion following
from the off-shell Goldstone superfield action and the one derived directly
from the nonlinear realizations formalism. We give a new derivation of the
off-shell Goldstone superfield actions for the considered systems, using a
universal procedure   inspired by the relationship between linear and 
nonlinear realizations of PBGS}.     
\end{abstract}

\vskip3cm
\begin{center}
{\it Submitted to Proceedings of the Seminar ``Classical and Quantum
Integrable Systems'' dedicated to the memory of M.V. Saveliev (Protvino,
Russia, 8--11 January 2001)} \end{center} 

\end{titlepage}
\section{Introduction}
One of the approaches to superbranes proceeds from the concept
of partial breaking of global  supersymmetry (PBGS) \cite{BW0}, 
\cite{HLP}-\cite{Kapu}. In such a description the objects
representing the physical worldvolume superbrane degrees of freedom  are
Goldstone superfields. The worldvolume  supersymmetry acts on them as linear
transformations and so is manifest. The rest of the full target supersymmetry
is realized nonlinearly. After passing to components in the Goldstone
superfield action and eliminating auxiliary fields, one recovers a
``static-gauge'' form of the appropriate Green-Schwarz-type action
(in general, after a field redefinition). 

While for the ordinary p-branes the worldvolume multiplets are scalar,
analogous supermultiplets  of Dp-branes are known to be vector, with the 
Born-Infeld dynamics for gauge fields (see \cite{Tseytlin} and
refs. therein). So the corresponding PBGS actions should form a subclass of
manifestly supersymmetric extensions \cite{CF}-\cite{KT} of the Born-Infeld
(BI) action. The actions from this variety are characterized by the second
nonlinearly realized hidden supersymmetry. The PBGS approach can  be
considered as an efficient tool for deducing such superextensions of the BI
action. Until now, only superextensions of abelian BI theory were derived  in
this way \cite{BG,IK1,BIK4,BIK5}. However, this approach could be useful 
in the non-abelian case too.\footnote{Its ``inborn'' feature (as distinct,
e.g., from the approach proceeding from gauge-fixed Green-Schwarz
Dp-brane actions \cite{APS}) is the manifestly {\it linear} realization of
unbroken supersymmetry.}   

Here we explain, on a few instructive examples ($N=1$ supermembrane,
space-filling D2- and D3-branes), how the  PBGS approach augmented with the
general methods of the theory of nonlinear realizations \cite{CWZ}
leads to a manifestly supersymmetric description of superbranes and
superextensions of the BI theory in terms of worldvolume superfields. What can
be directly deduced from the nonlinear realizations formalism in most cases,
is the Goldstone superfield dynamical equations  describing the given
superbrane or super BI theory on shell. The construction of the off-shell
superfield actions is more tricky, and it requires  constructing a {\it
linear} realization of the corresponding PBGS pattern. The superbrane or BI
superfield Lagrangian density is identified with a proper component  of some
linear supermultiplet of the full underlying supersymmetry. This multiplet is
subjected to covariant constraints which express all its components in terms
of the Goldstone multiplet of the unbroken  supersymmetry. The precise form of
these constraints can be found using the general relationship between linear
and nonlinear realizations of supersymmetries \cite{IvanovKapustnikov} 
adapted to the PBGS case in \cite{DIK,IKLZ}.\footnote{A different, though
seemingly equivalent adaptation was used in a recent preprint \cite{Kapu}.}
We apply this universal procedure to give a new derivation of the off-shell
Goldstone superfield actions for the examples considered. Also, in the
supermembrane case we prove the equivalence of the superfield equations of
motion derived from the off-shell action and those postulated in the pure
nonlinear realizations framework \cite{IK1,BIK3}.

\setcounter{equation}{0}
\section{N=1, D=4 supermembrane}
\noindent{\it 2.1   $N=1, D=4$ supermembrane as a PBGS system}. The
supermembrane in $D=4$ spontaneously breaks half of the $N=1,D=4$ 
supersymmetry and one translation. The set of generators of 
$N=1\; D=4$ Poincar\'e superalgebra in the $d=3$ notation is naturally
split into the unbroken $\left\{ Q_a, P_{ab} \right\}$ and  broken $\left\{
S_a, Z  \right\}$ parts ($a,b=1,2$). 
The basic anti-commutation relations in this notation read 
\be \left\{ 
Q_{a},Q_{b}\right\}=\left\{ S_{a},S_{b}\right\}=P_{ab},\; \left\{ 
Q_{a},S_{b}\right\} = \epsilon_{ab}Z \;.\label{susy3d} 
\ee 

As was argued in \cite{BIK3}, for deriving manifestly covariant superfield
equations describing the worldvolume dynamics of superbrane in the present
case (and some other ones), it suffices to deal with a
nonlinear realization of the superalgebra \p{susy3d}  itself, ignoring all 
generators of the automorphisms of \p{susy3d}. 
So we  put all generators  into the coset and associate  the $N=1\,,\, d=3$
superspace coordinates $\left\{  \theta^a, x^{ab} \right\}$ with $Q_a,
P_{ab}$. The remaining coset parameters are  Goldstone superfields, $\psi^a
\equiv \psi^a(x,\theta),\;q \equiv  q(x,\theta)$. A coset element $g$ is
defined by \footnote{In our notation the coset parameters  $x^{ab}$ and 
$q$ are imaginary, while $\theta^a$ and $\psi^a$ are real.} 
\be\label{coset3d}  g=e^{x^{ab}P_{ab}}e^{\theta^{a}Q_{a}}e^{qZ}   
e^{\psi^aS_a} \;.  \ee  Then one constructs the  Cartan 1-forms 
\bea 
&&g^{-1}d g =  \omega_Q^aQ_a + \omega_P^{ab} P_{ab} + \omega_ZZ + 
\omega_S^a S_a , \label{cartan13d} \\ 
&& 
\omega_P^{ab} =dx^{ab}+\frac{1}{4}\theta^{(a}d\theta^{b)} + 
  \frac{1}{4}\psi^{(a} d\psi^{b)}  \; ,\; \; \omega_Z  =  
dq+\psi_{a}d\theta^{a}, \;\;  \omega_Q^a =  d\theta^{a} ,\;\;
\omega_S^a=d\psi^{a}  \label{cartan3d} 
\eea 
and the corresponding covariant derivatives 
\be\label{cd3d} 
{\cal D}_{ab} =  (E^{-1})^{cd}_{ab}\,\partial_{cd} , \quad 
{\cal D}_a = D_a + \frac{1}{2}\psi^b D_a \psi^c \,{\cal D}_{bc}, 
\ee 
where 
\be 
E_{ab}^{cd}=\frac{1}{2}(\delta_a^c\delta_b^d+\delta_a^d\delta_b^c)+ 
  \frac{1}{4}(\psi^c\partial_{ab}\psi^d+
\psi^d\partial_{ab}\psi^c)~, \;\;D_a=\frac{\partial}{\partial \theta^a}+ 
\frac{1}{2}\theta^b\partial_{ab}, \;\;  \left\{ D_a, D_b \right\}
=\partial_{ab} \; . \label{flatd3d}  \ee 

The set of Goldstone superfields $\left\{
q(x,\theta),\psi^a(x,\theta)\right\}$  is reducible. Indeed,  $\psi_{a}$
appears  inside the form $\omega_Z$  {\it linearly} and so it can be
covariantly  eliminated by imposing the following manifestly covariant inverse
Higgs \cite{invh} constraint 
\be 
\left. \omega_Z\right|_{d\theta} = 0 \Rightarrow \psi_a={\cal D}_a q \;, 
\label{basconstr3d} 
\ee 
where $|_{d\theta}$ means the ordinary $d\theta$-projection of the form. 
Thus $q(x,\theta)$ is the only essential Goldstone superfield needed to 
describe the partial spontaneous breaking 
$N=1\,,\; D=4  \;\Rightarrow \; N=1\,,\; d=3$ within the coset scheme.   

In order to get dynamical equations, we put an additional, 
manifestly covariant constraint on the 
superfield $q(x,\theta)$. It is a direct covariantization of the
``flat'' equation of motion:
\be\label{eom13d} 
D^a D_a q=0 \quad \Rightarrow \quad  {\cal D}^a {\cal D}_a q=0 \;. 
\ee 
Eq. \p{eom13d} coincides with the dynamical
equation of the supermembrane in $D=4$ as it was given in
\cite{IK1}. It was derived  there from the coset approach with the $D=4$
Lorentz group generators included, so \p{eom13d} actually possesses the hidden
covariance under the full $D=4$ Lorentz group $SO(1,3)$. For the bosonic field
$q(x) \equiv q(x,\theta)|_{\theta=0}$ it yields the equation corresponding to
the static-gauge form of the Nambu-Goto action for membrane in
$D=4$.  Below we shall prove that \p{eom13d} is equivalent to the equation
following from an off-shell action of the supermembrane.       

Our next goal will be to give a new construction of this invariant
off-shell superfield action. As distinct from ref. \cite{IK1}, here we
apply the systematic approach based on the relationship between linear and
nonlinear  realizations of supersymmetry \cite{IvanovKapustnikov}. The
construction is quite similar to the one exploited in \cite{IKLZ} in
application to $d=2$ PBGS  systems. 

As a first step, we define a {\it linear} realization of the considered 
PBGS pattern $N=1, D=4 \rightarrow N=1, d=3$. From the $d=3$ point of
view, it amounts to $N=2 \rightarrow N=1$, with the $N=2, d=3$ 
Poincar\'e superalgebra given by the relations \p{susy3d}.
The primary object of such a realization is the scalar chiral $N=2, d=3$
superfield  $\Phi(x, \theta, \zeta)$, where $x^{ab}, \theta^a, \zeta^d$ are
the $N=2, d=3$ superspace coordinates. It is assumed to have the following
transformation property under the central charge operator $Z$: 
\be
Z\Phi = 1~. \label{Zrealiz}
\ee
This means that the central charge generator acts as shifts of $\Phi$. Such a
realization can be understood as the following specific coset realization of 
$N=2, d=3$ supersymmetry \p{susy3d}: one treats $\Phi$ as the coset parameter 
(Goldstone superfield) associated with $Z$, while the rest of coset parameters
as the coordinates of $N=2, d=3$ superspace on which $\Phi$  ``lives'' (cf. 
similar $d=2$ realizations considered in \cite{IKLZ}).
With respect to the $N=1$ supersymmetry $\{P_{ab}, Q_a\}$, the
superfield  $\Phi$ is a collection of standard $N=1$ superfields
(the coefficients in the expansion of $\Phi$ in $\zeta^a$), while  under the
$S$-supersymmetry it transforms in the  following way  
\be \delta_\eta \Phi =
-\eta^a\left( \frac{\partial}{\partial \zeta^a} -{1\over
2}\zeta^b\partial_{ab} - \theta_a Z \right) \Phi~. \label{Phitransf}  
\ee
Respectively, the spinor covariant derivatives in this realization are given
by 
\be
\hat{D}^\theta_a = \frac{\partial}{\partial \theta^a} +{1\over
2}\theta^b\partial_{ab} - \zeta_a Z = D_a - \zeta_a Z~, 
\quad  D^\zeta_a = \frac{\partial}{\partial \zeta^a} +{1\over
2}\zeta^b\partial_{ab}~. 
\ee
The covariant chirality condition reads 
\bea
\left(\hat{D}^\theta_a -i D^\zeta_a\right)\Phi = 0 \quad \Rightarrow
\quad \Phi = \phi - i \zeta^aD_a\phi + {1\over 4} \zeta^2
\left[ D^2\,\phi +2 i \right]~, \;\; \phi \equiv \phi(x, \theta)~,
\label{Phistruc}           
\eea
where \p{Zrealiz} was taken into account. Thus the complex $N=1$ superfield
$\phi(x,\theta)$ accommodates the irreducible set of the ($4+4$) off-shell
component fields of $\Phi(x,\theta,\zeta)$. Its $S$-supersymmetry 
transformation  directly stems from \p{Phitransf} and \p{Zrealiz}:  \be
\delta_\eta \phi = \eta^a\theta_a +i \eta^aD_a\phi~. 
\ee 
For the real superfields $\rho $ and $\phi_0$ defined by 
$$
\phi = \phi_0 + i\rho 
$$
we obtain the following transformation laws 
\be
\delta_\eta \rho = -i\eta^a\theta_a + \eta^aD_a\phi_0~, \quad
\delta_\eta\phi_0 = -\eta^aD_a\rho~. \label{rofi} 
\ee
The spinor superfield 
$$
\xi_a = iD_a\rho
$$     
transforms under the $S$-supersymmetry with an inhomogeneous shift 
\be
\delta_\eta \xi_a = \eta_a\left(1 -{i\over 2} D^2\phi_0\right) - {i\over 2}
\eta^b\partial_{ab}\phi_0~,  
\ee  
and so can be viewed as the Goldstone fermion of linear realization of the
same PBGS pattern $N=2 \rightarrow N=1, d=3$. The field content of
$\rho(x,\theta)$ coincides with that of $q(x,\theta)$, so $\rho$ can be
regarded as the $N=1$ Goldstone superfield for the spontaneously broken 
$Z$-transformations (it is shifted under $Z$). It is interesting to note that
the role of the inverse Higgs constraints in the linear realization is played
by the chirality condition \p{Phistruc} which expresses all $N=1$
superfield components of  $\Phi(x,\theta, \zeta)$ via derivatives 
of $\rho$ and $\phi_0$.    

Besides the basic Goldstone superfield $\rho$, there still remains the
superfield  $\phi_0$ possessing homogeneous transformation laws under both
$N=1, d=3$  supersymmetries. Now we shall show that it can be eliminated 
in terms of $\rho$ by imposing a nonlinear constraint which brings 
the considered linear realization into a nonlinear one
related  to the original nonlinear realization by a field redefinition. To
this end,  we apply the method of refs. \cite{IvanovKapustnikov},
\cite{DIK,IKLZ} to the system of $N=1$  superfields $\xi_a$, $\phi_0$.
Construct their {\it finite} $S$-supersymmetry transformation  and replace, in
the final expressions,  the parameters $\eta^a$ by the Goldstone superfields
$\psi^a(x,\theta)$ of the original nonlinear realization (taken with the sign
minus). The resulting objects  
\bea
&&\tilde \xi_a = \xi_a - \psi_a\left(1-{i\over 2}D^2\phi_0 \right) + {i\over
2} \psi^d\partial_{ad}\phi_0  -{1\over 4}\psi^2 \partial_{ab}\xi^b~, \nn     
&& \tilde \phi_0 = \phi_0 -i\psi^a\xi_a +{i\over 2} \psi^2 \left(1 - {i\over
2} D^2 \phi_0 \right) \label{sigfiel}   
\eea 
are homogeneously transformed under the $S$-supersymmetry (and under the $Q$
one, of course). So it is the covariant condition to put them equal to zero
\be
\tilde \xi_a = 0~, \quad \tilde \phi_0 = 0~. \label{sigmaEq}
\ee
Using the nilpotency property $\psi^3 =0$, it is easy to find that these
equations amount to 
\bea
\mbox{(a)}\;\; \psi^a = \frac{\xi^a}{1 -{i\over 2} D^2 \phi_0}~, \quad
\mbox{(b)}\;\;  \phi_0 ={i\over 2}\,\psi^2 \left(1-{i\over 2}D^2\phi_0
\right) =  {i\over 2}\,\frac{\xi^2}{1-{i\over 2}D^2\phi_0}~. \label{membrrel} 
\eea
These relations coincide (up to a slight difference in the notation) with  
those found in \cite{IK1}. The first one is the equivalence relation
between the nonlinear and linear realizations Goldstone fermions, while the
second one serves to express  $\phi_0$ in terms of $\psi^a$ or $\xi^a$: 
\be
\phi_0 = {i\over 2}\, \frac{\psi^2}{1 -{1\over 4} D^2 \psi^2} = 
\frac{i\,\xi^2}{1 +\sqrt{1+D^2\xi^2}}~. \label{lagrMem}
\ee
In view of the transformation property \p{rofi} of $\phi_0$, the integral 
\be
S \sim \int d^3xd^2\theta\, \phi_0 \label{membr1}
\ee
is invariant with respect to the whole $N=2, d=3$ supersymmetry, and so it is
the sought  off-shell action of the Goldstone superfield $\rho(x,\theta)$ (or
$q(x,\theta)$). It describes, in a manifestly worldvolume supersymmetric
manner, the $N=1, D=4$ supermembrane in a flat background. It can equally
be  written through the initial chiral $N=2$ superfield
$\Phi(x,\theta,\zeta)$, eq. \p{Phistruc}, as an integral over the full or
chiral $N=2, d=3$ superspaces 
\be 
S \sim \int d^3x d^2\theta d^2\zeta \Phi \bar\Phi \sim \int d^3x_Ld^2\theta_L
\Phi + \mbox{c.c.}~. \label{membr2}
\ee
Here $\theta_L^a = \theta^a - i\zeta^a$, $x^{ab}_L = x^{ab} +{i\over
2}\theta^{(a}\zeta^{b)}$, one should substitute into \p{Phistruc} the 
explicit expression for $\phi_0$ and, in the second case, pass to the
chiral basis in which  $\Phi$ does not depend on $\theta_R^a = \theta^a +
i\zeta^a $. In such a representation the full $N=2,\, d=3$ supersymmetry
\p{susy3d}  is manifest. Note that the two invariants \p{membr2} are
independent before  passing to the nonlinear realization,  but they 
become identical (up to a numerical factor) in the nonlinear realization
framework. A similar phenomenon takes place in other PBGS cases
\cite{RT,Tseytlin,IKLZ}.            

Note that the same relations \p{membrrel} could be obtained by imposing,
in the spirit of \cite{RT}, the  nilpotency condition on the
appropriate real $N=2$ superfield constructed out of $\Phi$ and its conjugate
and having a zero central charge \cite{BZ,Kapu}. Our purpose here was to
demonstrate that the generic method used earlier in  refs \cite{DIK,IKLZ}
works fairly well in this case too.  \vspace{0.2cm}

\noindent{\it 2.2  Equivalence of two forms of the supermembrane
equations of motion.} Here we show that the dynamical equation \p{eom13d}
postulated on the purely geometric grounds and the equation of motion
following from the  off-shell action \p{membr1} are equivalent to each other.       

Eq. \p{eom13d} written in terms of the Goldstone fermion superfield $\psi_a$  
\be
{\cal D}^a\psi_a = 0~, 
\ee
can be cast in the following more detailed form 
\be \label{I} W +{1\over 2} \psi^a W^{bc}\partial_{ab}\psi_c -{1\over
16}\psi^2  W^{ac}\partial_{bc}\psi_d\partial^{bd}\psi_a = 0~, 
\ee
where   
\be \label{def1}
W \equiv D^a\psi_a~, \quad W^{ab} \equiv D^{(a}\psi^{b)}
\ee
and the explicit expression \p{cd3d} for the covariant derivative ${\cal D}_a$ 
was used. In such a form the equation includes only flat derivatives
$D_a$ and $\partial_{ab}$.

On the other hand, the equation of motion which follows from the
off-shell  action \p{membr1} by varying it with respect to the unconstrained
$N=1$ superfield $\rho$ can be written in terms of $\psi^a$   
\be \label{II}
D^a\lambda_a = 0~, 
\ee
where 
\be 
\lambda_a \equiv  {\psi_a \over 1 - {1\over 2}D\psi\cdot D\psi} 
+{1\over 4}{\psi^2 \over (1 - {1\over 2}D\psi\cdot D\psi)^2} \left\{ 
\partial_{ab}\psi^b +2D_b\psi_a D_d\psi_c \partial^{db}\psi^c \right\}
\label{def2}
\ee 
and $A\cdot B \equiv  A^{ab}B_{ab}$. This form of the equation can be
deduced from the equivalence relation \p{membrrel}. Let us start from the
Lagrangian \p{lagrMem} in the form   
\be \label{action}
\phi_0  = {i\over 2}{\xi^2 \over 1  - {i\over 2}D^2\phi_0} = {i\over 2} \psi^2
(1- {i\over 2} D^2\phi_0)~,  \quad S \sim \int d^3 x d^2 \theta \;\phi_0(x,
\theta) \ee
with 
\be
(1 -{i\over 2} D^2\phi_0)_{eff} = {1\over 1 +{1\over 2}D\psi\cdot D\psi}~.
\ee 
Here the subscript ``{\it eff}'' means that we ignore all terms $\sim
\psi^a$ (but not those with derivatives on $\psi^a$). 
The variation of $\phi_0$ reads  
\be \label{var} 
\delta\phi_0 = i \delta\xi^a\psi_a -  {1\over 4} \psi^2
(D^2\delta\phi_0)_{eff}~, 
\ee
where $(D^2\delta\phi_0)_{eff}$ is determined from the equation 
\be
(D^2\delta\phi_0)_{eff}\left( 1 - {1\over 2} D\psi\cdot D\psi \right) 
= -2i\, D^l\delta \xi^a D_l\psi_a + i \delta\xi^a D^2\psi_a~.
\ee
Using eq. \p{membrrel} and the kinematical constraint 
\be \label{constr}
D^2\xi_a =\partial_{ab}\xi^b
\ee
following from the relation $\xi_a = iD_a\rho$, one obtains, in zeroth
order in $\psi$,
\be  \label{kin}
D^2\psi_a = \partial_{ab}\psi^b + D_b\psi_a D_d\psi_c\partial^{bd}\psi^c 
+ O(\psi)~.
\ee
After substituting all this in \p{var} and integrating in $\delta_\eta S$ by
parts with  keeping in mind that $\delta \xi^a = iD^a \delta \rho$, one
obtains  the equation of motion in the form \p{II}, \p{def2}. 

Let us prove the equivalence of eqs. \p{I} and \p{II}. First,  
we shall show that \p{II} is identically satisfied if \p{I} holds.

As a preparatory step, we extract some corollaries of \p{I}. Acting
on \p{I} by  spinor derivative, one gets, in the zeroth order in $\psi$
\be \label{cor1}
D^2\psi_a = - \partial_{ab}\psi^b - W_{ab} W_{cd}\partial^{bd}\psi^c 
+ O(\psi)~.
\ee
Comparing it with the kinematical constraint \p{kin}, one finds  
\be  \label{dyn}
\mbox{(a)} \;\;D^2\psi_a = O(\psi)~, \quad
\mbox{(b)}\;\;\partial_{ab}\psi^b + W_{ab} W_{cd}\partial^{bd}\psi^c= O(\psi)~.
\ee
Further, in the same order, acting on \p{I} by $D^2$ ($D^2D^a = -
\partial^{ab}D_b$), one obtains
\be  \label{cor2}
\partial^{ab}W_{ab} = - W^{ab} W^{dc} \partial_{bc} W_{ad} + O(\psi)~,
\ee
where eq.(\ref{dyn}b) was used. 

Let us apply to eq. \p{II}. Firstly, owing to eq.(\ref{dyn}b) one can
make in \p{II} the substitution    
\be \label{repl1}
\lambda_a \; \Rightarrow \;
{\psi_a \over 1 - {1\over 2}D\psi\cdot D\psi} 
- {1\over 4}{\psi^2 \over (1 - {1\over 2}D\psi\cdot D\psi)^2}\; 
\partial_{ab}\psi^b~.
\ee 
Secondly, because of the relation $W^2 \sim \psi^2$ following from \p{I}, 
one can replace altogether  
\be \label{repl2}
(D\psi\cdot D\psi) \;\Rightarrow \; (W\cdot W)~.
\ee
After expressing $W$ from \p{I} and using eq. (\ref{dyn}a) together with the
identity  \be \label{iden}
\partial_{ab}\psi^b\partial^{ad}\psi^c W_{dc} = O(\psi)
\ee
which follows from (\ref{dyn}b), eq. \p{II} can be rewritten, up to  
the non-singular factor 
$$
1 - {1\over 2} (W\cdot W) \equiv F, 
$$
in the following equivalent form 
\bea 
&& (W\cdot W)\, \psi^a \partial_{ab}\psi_c\, W^{bc}  + \psi^2\, 
W^{ab}W^{dc}\partial_{bc}W_{ad} - 2 (G + K) = 0~,  \label{IIfin} \\
&& G \equiv \psi_a\,W^{ab} D^2\psi_b\, ~, \quad K \equiv 
\psi_a\,W^{ab}\,\partial_{bc}\psi^c~. \label{def3}
\eea
It remains to compute $G$ and $K$. For this one should know 
$D^2\psi^a$ and $\partial_{ab}\psi^b$ up to the terms $\sim \psi$, while 
eqs. \p{dyn} define them only up to the terms $\sim 1$. To find these  
quantities, one should evaluate $D^2\psi^a$ at the required order both from
the kinematical  constraint (using eqs. \p{membrrel} and \p{constr}) 
and from \p{I} and compare these two expressions. They can be
straightforwardly  computed, but look rather cumbersome. Much simpler are 
the corresponding expressions for $G$ and $K$. Starting from 
the kinematical eqs. \p{constr}, \p{membrrel}, expressing $W$ from \p{I}
and using  the dynamical identity \p{iden} combined with cyclic identities, 
one finds 
\be \label{kin2}
G = K  - {1\over 2} (W\cdot W) 
\,\psi^a\,W^{bc}\,\partial_{ab}\psi_c +{1\over 2} 
\psi^2\, W^{ab}\,W^{cd}\,\partial_{ac}W_{bd}~.
\ee
On the other hand, the direct calculation making use only of eq. \p{I} 
and its corollaries yields 
\be \label{dyn2}
G = -K  + 
{1\over 2} (W\cdot W) \,\psi^a\,W^{bc}\,\partial_{ab}\psi_c +{1\over 2} 
\psi^2\, W^{ab}\,W^{cd}\,\partial_{ac}W_{bd}~.
\ee
Comparing these expressions, one finds 
\bea \label{GK}
G = {1\over 2} \psi^2\, W^{ab}\,W^{cd}\,\partial_{ac}W_{bd}~, 
\quad K = {1\over 2} (W\cdot W) \,\psi^a\,W^{bc}\,\partial_{ab}\psi_c~. 
\eea
Substituting them into eq. \p{IIfin}, one finds the latter to be identically 
satisfied. Thus we have shown that eq. \p{II} is fulfilled as a consequence 
of eq. \p{I}. 

To prove the equivalence of eqs. \p{I}, \p{II}, it remains to show that 
eq. \p{II} necessarily implies \p{I}. As a first step, one observes that
\p{II}  implies the same relations \p{cor1}, \p{cor2}, \p{dyn}, 
\p{iden} as eq. \p{I} and so we are allowed to use them in the course of the 
proof. In particular, using in addition the fact that $W\sim \psi, \psi^2$ 
as another corollary of \p{II}, one can make the replacements \p{repl1}, 
\p{repl2} in \p{II}. Further, from the kinematical constraints and \p{II} 
one derives the {\it same} expressions  \p{GK} for $G$ and $K$ 
(obtained earlier starting from \p{I}). Substituting them into \p{II}, 
one gets 
\be 
W +{1\over 2} \psi^a W^{bc}\partial_{ab}\psi_c = 0
\ee
that actually amounts to eq. \p{I} due to the relation (\ref{dyn}b) 
and its corollary \p{iden}. 

Thus we have proven the equivalence of eq. \p{I} derived 
from the purely geometric setting and eq. \p{II} obtained from 
the variation  principle associated with the off-shell action  
\p{action}. An important consequence of this  
fact is that the action \p{action} and eq. \p{II} possess all symmetries 
encoded in \p{I}, including the $D=4$ Lorentz symmetry.

\setcounter{equation}{0}
\section{Space-filling D2-brane}
As the second instructive example, we consider the ``space-filling'' 
D2-brane with the $N=1,\,d=3$ vector multiplet as the worldvolume one. 
\vspace{0.2cm}

\noindent{\it 3.1  D2-brane dynamics from nonlinear realizations}. Our
starting point is the superalgebra \p{susy3d} with $Z=0$. The coset element
$g$ contains only one Goldstone superfield $\psi^a$, and the covariant
derivatives  are still given by \p{cd3d}. 
In the flat case the $d=3$ vector multiplet is described  by a $N=1$ spinor
superfield strength $\mu_a$ subjected to the Bianchi identity: 
\be\label{cc13d} 
D^a\mu_a=0 \; \Rightarrow \; \left\{ 
 \begin{array}{l} 
   D^2 \mu_a=-\partial_{ab}\mu^b~,  \nn 
   \partial_{ab}D^a\mu^b = 0~ . \nonumber 
  \end{array} \right. 
\ee 
Its equation of motion reads 
\be\label{flateom23d} 
D^2 \mu_a =0 \; . 
\ee 
It was shown in \cite{BIK3} that the following manifestly covariant
generalization of \p{cc13d}, \p{flateom23d}  describes the D2-brane: 
\be\label{eom23d}   \mbox{(a)}\;\;{\cal D}^a\psi_a=0~, \quad \mbox{(b)}\; \;
{\cal D}^2 \psi_a =0~.  \ee 
The reasoning was mainly based on the observation that the purely bosonic
limit of \p{eom23d} amounts to the following  equation for the vector
$V_{ab}\equiv {\cal D}_a\psi_b|_{\theta=0}$:  
\be\label{boseq13d} 
\left( \partial_{ac} +V_a^d V_c^f \partial_{df}\right) V_b^c =0 \;. 
\ee 
This nonlinear but polynomial equation was shown to be a
``disguised'' form of the equations of the non-polynomial $d=3$ BI action
which is just the bosonic core of the superfield D2-brane PBGS action as
was explicitly demonstrated in \cite{IK1}. The passing to the standard form 
of the $d=3$ BI equation is achieved by a field redefinition which is a
bosonic limit of the superfield equivalence redefinition relating the
nonlinear realization Goldstone fermion $\psi_a$ to $\mu_a$ treated as 
the Goldstone fermion of a {\it linear} realization of the same PBGS pattern
(see next Subsection). Using this equivalence, one may explicitly show, like
in the supermembrane case, that the equations \p{eom23d}
are equivalent to the worldvolume superfield equation  following from the
off-shell D2-brane action given in \cite{IK1}.    
\vspace{0.2cm}

\noindent{\it 3.2  Off-shell superfield D2-brane action}. Now we shall
re-derive the off-shell D2-action of ref. \cite{IK1} by the same generic
method which was applied above to construct the Goldstone superfield action of
$N=1, D=4$ supermembrane. To define the  appropriate linear realization of the
considered PBGS pattern, one needs  to embed the $N=1, d=3$ Maxwell superfield
strength $\mu_a$ into a linear $N=2, d=3$ multiplet. The latter should have
such a transformation law under the $S$-supersymmetry that  $\mu_a$ transform
with an inhomogeneous term $\sim \eta_a$ and so admit an interpretation as the
Goldstone fermion of  linear realization. 

The appropriate $N=2, d=3$ supermultiplet was proposed in \cite{BZ} as a
deformation of the $N=2, d=3$ Maxwell multiplet (which is
a dimensional reduction  of the $N=1, d=4$ tensor multiplet). In our notation
this deformed multiplet  is described by a real $N=2, d=3$ superfield
$W(x,\theta, \zeta)$  subjected to the following constraints   \be
\mbox{(a)}\;\;\left[ (D)^2 - (D^\zeta)^2 \right] W = -2i~, \quad  
\mbox{(b)} \;\; D^{a}D^{\zeta}_a W = 0 \label{constrW}
\ee
(this form of constraints is most convenient for our purposes, it can
be obtained from the one given in \cite{BZ} by choosing a  specific
frame with respect to  the explicitly broken $U(1)$-automorphism symmetry and
making an appropriate rescaling of $W$ \footnote{For the first
time such a deformation of the  $N=1, d=4$ tensor multiplet constraints was
considered in \cite{IKL} in the context of $N=4$ superconformal
mechanics.}). 

The standard $S$-supersymmetry transformation law of $W$  
\be    
\delta_\eta W = -\eta^a\left( \frac{\partial}{\partial \zeta^a} -{1\over
2}\zeta^b\partial_{ab} \right) W  \label{Wtrnas}
\ee
implies the following transformation laws for the irreducible 
$N=1$ superfield components of $W(x,\theta, \zeta)$, $\mu_a \equiv
-iD^\zeta_a W|_{\zeta=0}$ and $w \equiv W|_{\zeta=0}$,   
\bea
\mbox{(a)} \;\;\delta_\eta \mu_a = \eta_a\left(1- {i\over 2} D^2 w\right) +
{i\over 2}\eta^b\partial_{ab}w~, \quad \mbox{(b)}\;\;\delta_\eta w = -i
\eta^a\mu_a~. \label{Wtrans2} 
\eea  
It is easy to check that eq. (\ref{Wtrans2}a) is consistent with the Bianchi 
identity \p{cc13d} (which is none other than eq. (\ref{constrW}b)). Just due to
the presence of  constant $U(1)_A$ breaking term in the r.h.s. of
(\ref{constrW}a), the $N=1$ Maxwell superfield  $\mu_a$ transforms
inhomogeneously under the $S$-supersymmetry,  and thus is recognized as the
Goldstone fermion of the linear realization  of the considered $N=2
\rightarrow N=1, d=3$ PBGS pattern.

Like in the supermembrane case, the additional homogeneously transforming 
$N=1$ superfield $w(x,\theta)$ can be traded for the Goldstone-Maxwell 
one $\mu_a$ by imposing nonlinear constraints the precise form of which
is dictated by our generic method applied to the given system. 
As the first step, one defines the superfields $\tilde{\mu}_a$ and
$\tilde{w}$ as finite  $S$-supersymmetry transforms of $\mu_a$ and $w$, with
the supertranslation parameter $\eta^a$ being replaced by $-\psi^a(x,\theta)$
\bea
\tilde{\mu}_a = \mu_a - \psi_a\left(1- {i\over 2} D^2 w\right) -{i\over 2}
\psi^b\partial_{ab}w -{1\over 4} \psi^2 \partial_{ab}\mu^b ~, \;
\tilde{w} = w + i\psi^a\mu_a - {i\over 2}\,\psi^2 \left(1- {i\over 2}
D^2w\right). 
\eea
These quantities homogeneously transform under all $N=2, d=3$
transformations and so one can covariantly equate them to
zero  
\be
\tilde{\mu}_a = \tilde{w} = 0~.
\ee
From these covariant constraints one gets the equivalence relation between
$\psi^a$  and $\mu^a$ 
\be
\psi^a = \frac{\mu^a}{1- {i\over 2} D^2 w}~, 
\ee 
as well as  the relation 
\be
w = -{i\over 2}\,\frac{\mu^2}{1- {i\over 2} D^2 w}~.
\ee 
These are precisely the equations derived in \cite{IK1} (up to a rescaling of
$w$). They can be used to  express $w$ in terms of either $\psi^a$, or $\mu^a$ 
\be 
w = -{i\over 2}\, \frac{\psi^2}{1 + {1\over 4} D^2 \psi^2} =
-\frac{i\,\mu^2}{1 + \sqrt{1 - D^2\mu^2}}~. \label{D2dens}
\ee         
This composite superfield is just the corresponding Goldstone superfield
Lagrangian density, 
\be
S \sim \int d^3x d^2\theta\, w~, \label{muaction}
\ee
since, in virtue of the Bianchi identity \p{cc13d}, the $d^3xd^2\theta$
integral  of the variation (\ref{Wtrans2}b) is vanishing, i.e. $ \delta_\eta
S = 0$.

The same superfield D2-brane action can be written in a manifestly
$N=2$ supersymmetric  form as an integral over the whole $N=2$ superspace,
with either $W^2$ or the $N=2, d=3$ Fayet-Iliopoulos term as the Lagrangian
densities ( like in other PBGS cases, these two independent invariants are
reduced to each other after passing to the nonlinear realization). 
     
\section{Space-filling D3-brane}
As the last example we consider the space-filling D3-brane in $d=4$. 
This system amounts to the PBGS pattern 
$N=2 \rightarrow  N=1$ in $d=4$, with a nonlinear generalization of 
$N=1, \, d=4$ vector multiplet as the Goldstone multiplet
\cite{BG,RT}. The  off-shell superfield action for
this system was constructed in \cite{CF,BG}. Here we explain, 
following ref. \cite{BIK3},  how  the corresponding dynamical equations can be
derived directly from the coset approach, like in other cases considered in
this paper. As a new reuslt, we shall recover the action of \cite{BG} by
means of the universal procedure exemplified in the previous Sections, thus
establishing a direct link between the approach of refs. \cite{BG,RT} 
and the customary coset approach. 
\vspace{0.2cm}

\noindent{\it 4.1  D3-brane superfield equations of motion from nonlinear
realizations}. Our starting point is the $N=2,\, d=4$ Poincar\'e superalgebra
without  central charges: 
\be 
\left\{ Q_{\alpha}, {\bar Q}_{\dot\alpha} \right\}=2P_{\alpha\dot\alpha} 
\;,\; 
\left\{ S_{\alpha}, {\bar S}_{\dot\alpha} \right\}=2P_{\alpha\dot\alpha} 
\;.\label{susyD43d} 
\ee 
Assuming the $S_{\alpha}, {\bar S}_{\dot\alpha}$ supersymmetries 
to be spontaneously broken, we introduce the Goldstone superfields 
$\psi^{\alpha}(x,\theta,\bar\theta), \, 
{\bar\psi}^{\dot\alpha}(x,\theta,\bar\theta)$ as the corresponding parameters 
in the following coset 
\be 
g=e^{ix^{\alpha\dot\alpha}P_{\alpha\dot\alpha}} 
   e^{ i\theta^{\alpha}Q_{\alpha}+ 
   i{\bar\theta}_{\dot\alpha}{\bar Q}^{\dot\alpha}} 
e^{ i\psi^{\alpha}S_{\alpha}+ 
   i{\bar\psi}_{\dot\alpha}{\bar S}^{\dot\alpha}} \;. 
\ee 
With the help of the corresponding Cartan forms one can define the covariant
derivatives  \bea\label{cd43d} 
{\cal D}_{\alpha}=  D_{\alpha}-i\left( 
{\bar\psi}^{\dot\beta}D_{\alpha}\psi^{\beta} + 
  \psi^{\beta}D_{\alpha}{\bar\psi}^{\dot\beta}\right) {\cal 
   D}_{\beta\dot\beta},  \quad 
{\cal D}_{\alpha\dot\alpha}= 
 \left( 
E^{-1}\right)_{\alpha\dot\alpha}^{\beta\dot\beta}\partial_{\beta\dot\beta} 
  \;, 
\eea 
where 
\bea 
D_{\alpha}=\frac{\partial}{\partial\theta^{\alpha}} - 
i{\bar\theta}^{\dot\alpha}\partial_{\alpha\dot\alpha}~, \;\; 
{\bar D}_{\dot\alpha}= 
 -\frac{\partial}{\partial{\bar\theta}^{\dot\alpha}} + 
i{\theta}^{\alpha}\partial_{\alpha\dot\alpha}~,\;\;
E_{\alpha\dot\alpha}^{\beta\dot\beta}= \delta_{\alpha}^{\beta}  
\delta_{\dot\alpha}^{\dot\beta} 
-i\psi^{\beta}\partial_{\alpha\dot\alpha}{\bar\psi}^{\dot\beta}- 
i{\bar\psi}^{\dot\beta}\partial_{\alpha\dot\alpha}\psi^{\beta}\;. 
\label{flat4d3d}  \eea 

Now we can write the covariant version of the constraints 
on $\psi^{\alpha},\,{\bar\psi}^{\dot\alpha}$ which define the superbrane 
generalization of $N=1,\,d=4$ vector multiplet, together with the covariant 
equations of motion for this system. They are a direct covariantization of 
the free $N=1,\, d=4$ Maxwell superfield strength constraints and 
equation of motion: 
\be
\label{chir5e3d} 
\mbox{(a)}\;\; {\overline{\cal D}}_{\dot\alpha}\psi_{\alpha}=0~, \;\; {\cal 
D}_{\alpha}\bar\psi_{\dot\alpha}=0~, \qquad  
\mbox{(b)} \;\; {\cal D}^{\alpha}\psi_{\alpha}=0\;,\;\;  {\overline{\cal 
D}}_{\dot\alpha}{\overline \psi}^{\dot\alpha}=0 \;. 
\ee 
Eqs. (\ref{chir5e3d}a) are a covariantization of the flat $N=1$ chirality
conditions while (\ref{chir5e3d}b) generalizes at once the $N=1$
superfield strength Bianchi identity and equation of motion. 
As was argued in \cite{BIK3}, this  set of superfield equations 
is self-consistent and compatible with the algebra of
the  covariant  derivatives \p{cd43d}. For the physical bosonic components
of $\psi,\bar\psi$,  
\be\label{defcomp53d} 
V^{\alpha\beta}\equiv {\cal 
D}^{\alpha}\psi^{\beta}|_{\theta=0}~, \quad {\bar V}^{{\dot\alpha}\dot\beta} 
\equiv   {\overline{\cal 
D}}^{\dot\alpha}{\bar\psi}^{\dot\beta}|_{\theta=0}, 
\ee 
these superfield equations imply, in the purely bosonic limit, the following
equations   
\bea\label{eom63d} 
\partial_{\alpha\dot\alpha}V^{\alpha\beta}-V_{\alpha}^{\gamma} 
 {\bar V}_{\dot\alpha}^{\dot\gamma}\;\partial_{\gamma\dot\gamma} 
  V^{\alpha\beta} =0~,\quad 
\partial_{\alpha\dot\alpha}{\bar 
V}^{{\dot\alpha}\dot\beta}-V_{\alpha}^{\gamma} 
 {\bar V}_{\dot\alpha}^{\dot\gamma}\;\partial_{\gamma\dot\gamma} 
  {\bar V}^{{\dot\alpha}\dot\beta} =0 \;. 
\eea 
It was shown in \cite{BIK3} that, like the analogous equations \p{boseq13d} in 
the D2-brane case, these equations can be cast in the standard form 
of the $d=4$ BI theory equations augmented with the Bianchi identity for 
the Maxwell field strength. 

Note that at the full superfield level the field redefinition which
leads from the disguised form of the BI equations \p{eom63d} to their
``canoical'' form corresponds to passing from the Goldstone fermions 
$\psi_\alpha$,  $\bar\psi_{\dot\alpha}$  
to the standard Maxwell superfield  strength $W_\alpha, \bar W_{\dot\alpha}$.
The nonlinear action of \cite{CF,BG,RT}  was written just in terms of
this latter  object. The equivalent form \p{chir5e3d} of the equations of 
motion and  Bianchi identity is advantageous in that it manifests the second
(hidden) supersymmetry, being constructed out of  the  covariant objects. 
\vspace{0.2cm}

\noindent{\it 4.2  Linear and nonlinear realizations of the $N=2 \rightarrow
N=1$ PBGS}. Now we wish to precisely establish the correspondence just
mentioned and to reproduce the off-shell BI action of \cite{CF,BG,RT} by
applying the general techniques based on the relationship between linear and
nonlinear realizations of PBGS, like in the previous Sections. 

Our starting point is the $N=2, d=4$ Goldstone-Maxwell multiplet
\cite{APT,BG,IZ}. In the $N=2$ superspace $(x^{\alpha\dot\alpha},
\theta^\alpha_i, \bar\theta^{\dot\alpha i})$ it is defined by the following 
deformation \cite{IZ} of the standard $N=2$ Maxwell superfield strength
constraints   
\be
\mbox{(a)} \;\;D^{ik}W -\bar{D}^{ik}\bar W = i M^{(ik)}~, \qquad
\mbox{(b)} \;\;D^i_\alpha \bar W = \bar D_{\dot\alpha i} W =0~.
\label{deformN2}  
\ee
Here
$$
D^i_\alpha  = \frac{\partial}{\partial \theta^\alpha_i} -i
\bar\theta^{\dot\alpha i}\partial_{\alpha\dot\alpha}~, \;
\bar D_{\dot\alpha i} = -\frac{\partial}{\partial \bar\theta^{\dot\alpha i}}
+i \theta^{\alpha}_i \partial_{\alpha\dot\alpha}~, \; 
D^{ik} = D^{\alpha i}D^k_\alpha~, \; \bar D^{ik} = \bar D_{\dot\alpha}^i \bar
D^{\dot\alpha k} 
$$ 
and $M^{ik} = M^{ki}$ is a triplet of constants which
explicitly break the automorphism  $SU(2)_A$ of $N=2$ supersymmetry down to
$U(1)_A$ and satisfy the pseudo-reality  condition 
$$ 
\overline{(M^{ik})} = \epsilon_{in}\epsilon_{km}M^{nm}~.
$$
In components, the deformation (\ref{deformN2}a) amounts to the appearance of
constant imaginary part $\sim M^{ik}$ in the isotriplet auxiliary field of
$N=2$ Maxwell multiplet. 

Now we pass to the $N=1$ superfield notation by relabelling the Grassmann
coordinates and spinor derivatives as  
$$
\theta^\alpha_1 \equiv \theta^\alpha~, \;\theta^\alpha_2 \equiv \zeta^\alpha,
\; D^1_\alpha \equiv D_\alpha, \; D^2_\alpha \equiv D^\zeta_\alpha~. 
$$
In order to have the off-shell $S$-supersymmetry (acting as
$\zeta$-supertranslations) spontaneously broken while the $Q$-supersymmetry
unbroken, we are led to choose the following frame with respect to the
explicitly broken $SU(2)_A$ \be
M^{12} = 0~, \quad M^{11} = M^{22} = m~, 
\ee
where $m$ is a real constant. Like in the case of D2-brane it is fixed up to 
rescaling of  $W$. A convenient choice is 
$$
m = -2~. 
$$
It will be also convenient to choose the basis in $N=2$ superspace where the
chirality with respect to the variable $\zeta^\alpha$ is manifest
\be
\bar D^\zeta_{\dot\alpha} = -\frac{\partial}{\partial
\bar\zeta^{\dot\alpha}}~, \quad   D^\zeta_{\alpha} = \frac{\partial}{\partial
\zeta^{\alpha}} -2i \bar\zeta^{\dot\alpha}\partial_{\alpha\dot\alpha}~.
\ee
In this basis, constraints \p{deformN2} imply the following structure of
the superfield $W(x, \theta, \zeta)$
\be
W = i\phi + i\zeta^\alpha W_\alpha -i {1\over 2} \zeta^2\left(1 +{1\over 2}
\bar D^2 \bar\phi\right)~, \label{N2MG}
\ee
where $\phi$ and $W_\alpha$ are chiral $N=1$ superfields   
\be
\bar D_{\dot\alpha} \phi = \bar D_{\dot\alpha} W_\alpha = 0~, 
\ee
and the fermionic superfield $W_\alpha$ obeys the $N=1$
Maxwell superfield strength constraint 
\be 
D^\alpha W_\alpha + \bar D_{\dot\alpha}\bar W^{\dot\alpha} = 0~.
\ee
The numerical factors in \p{N2MG} were chosen for further convenience.

The $S$-supersymmetry transformation of the $N=2$ superfield $W$ 
\be 
\delta_\eta W = -\left[\eta^\alpha \frac{\partial}{\partial
\zeta^\alpha} + \bar\eta^{\dot\alpha}\left(\frac{\partial}{\partial
\bar\zeta^{\dot\alpha}} +2i \zeta^\alpha
\partial_{\alpha\dot\alpha}\right)\right] W
\ee
implies the following ones for its $N=1$ superfield components $\phi$ and
$W_\alpha$  
\bea
&& \delta_\eta \phi = -(\eta W)~, \quad \delta_\eta \bar\phi = -(\bar
W\bar\eta)~, \nn      
&&\delta_\eta W_\alpha = \eta_\alpha \left(1+{1\over 2} \bar D^2 \bar \phi
\right) + 2i \bar\eta^{\dot\alpha} \partial_{\alpha\dot\alpha}\phi~, \quad 
\delta_\eta \bar W_{\dot\alpha} = \overline{(\delta_\eta W_\alpha)}~.
\label{tranfiW}
\eea
The superfield $W_\alpha$ shows up an inhomogeneous shift $\sim \eta_\alpha$
(proportional to the $SU(2)_A$ breaking parameters) in its transformation, so
it is the Goldstone fermion of the  linear realization of the considered $N=2
\rightarrow N=1, \;d=4$ PBGS pattern (the Goldstone-Maxwell $N=1$ superfield). 

Now we are prepared to start the algorithmic procedure of
passing to the relevant nonlinear realization exemplified in the previous
Sections.  We construct the finite $\eta$-transformations of the superfields
$\phi$  and $W_\alpha$ proceeding from the infinitesimal ones \p{tranfiW}
\be
\{\,\phi (\eta)~,\;\; W_\alpha(\eta) \,\} 
= \left( 1 + \delta_\eta + {1\over
2} \delta^2_\eta +{1\over 3!}\delta^3_\eta + {1\over 4!}\delta^4_\eta \right) 
\{\,\phi~, \;\;W_\alpha \,\}~, 
\ee
then pull out the parameters $\eta_\alpha, \bar\eta_{\dot\alpha}$ to the left
and replace them by   the original nonlinear realization Goldstone fermions,
$\eta_\alpha \rightarrow -\psi_\alpha, \;\bar\eta_{\dot\alpha}
\rightarrow -\bar\psi_{\dot\alpha}$. It is a matter of
straightforward  computation to check that the objects $\tilde \phi
\equiv \phi(-\psi), \tilde W_\alpha \equiv W_\alpha (-\psi)$  transform
homogeneously (though nonlinearly) with respect to the $\eta$-transformations
\be \delta_\eta \{\,\tilde\phi, \; \tilde W_\alpha \,\} =
i\left(\psi^\alpha\bar\eta^{\dot\alpha} - \eta^\alpha
\bar\psi^{\dot\alpha}\right)\partial_{\alpha\dot\alpha} \{\, \tilde\phi,
\; \tilde W_\alpha\, \}~, 
\ee
and behave as ordinary $N=1$ superfields under the unbroken
$\epsilon$-supertranslations acting in the $N=1$ superspace $(x, \theta,
\bar\theta)$. Hence, one can impose the covariant  constraints
\be
\tilde\phi  = \tilde W_\alpha = 0~. \label{basicD3}
\ee
Explicitly, the relations between $\phi, W_\alpha$ and $\psi_\alpha$
implied by these constraints are as follows 
\bea 
\phi &=& -{1\over 2}\psi^2 \left(1 +{1\over 2}\bar D^2 \bar \phi\right) -
i\psi^\alpha\bar\psi^{\dot\alpha} \partial_{\alpha\dot\alpha}\phi - i\psi^2
\bar\psi_{\dot\alpha}\partial^{\dot\alpha\beta}W_\beta -{3\over 8} \psi^4
\Box \phi~, \label{phiExpr} \\
W_\alpha &=& \psi_\alpha \left(1 +{1\over 2}\bar D^2 \bar \phi\right) +2i
\bar\psi^{\dot\alpha}\partial_{\alpha\dot\alpha}\phi +i \left(\psi_\alpha
\bar\psi^{\dot\alpha}\partial_{\beta\dot\alpha}W^\beta +
\psi_\beta\bar\psi^{\dot\alpha}\partial_{\alpha\dot\alpha}W^\beta \right) \nn 
&& + \, {1\over 2}\left( \bar\psi^2 \psi_\alpha \Box \phi -{i\over 2} \psi^2
\bar\psi^{\dot\alpha} \partial_{\alpha\dot\alpha}\bar D^2 \bar \phi \right) -
{1\over 8} \psi^4 \Box W_\alpha~, \quad \Box \equiv
\partial_{\alpha\dot\alpha}\partial^{\dot\alpha\alpha}~. \label{Wexpr}
\eea                

These equations look a bit more complicated as compared to the
previous examples, but, nevertheless, they can be treated in 
the precisely same algorithmic way. One should firstly make use of 
the relation \p{Wexpr} (and its conjugate) to express $\psi_\alpha$,
$\bar\psi$ in terms of $W_\alpha$, $\bar W_{\dot\alpha}$ and their
$x$-derivatives, and then substitute these expressions  into \p{phiExpr} and
its conjugate, thus obtaining covariant relations between $\phi, \bar \phi$
and $W_\alpha, \bar W_{\dot\alpha}$. The latter should allow one to trade
$\phi, \bar\phi$ for $W_{\alpha}, \bar W_{\dot\alpha}$ (or  for $\psi_\alpha,
\bar\psi_{\dot\alpha}$ in view of the equivalence relation between these two
kinds of the Goldstone fermion). A technically more simple way  to arrive at
the final relations is as follows.  One computes $W^2$ from \p{Wexpr} and in
the obtained relation  $$
W^2 = \psi^2 \left(1 +{1\over 2}\bar D^2 \bar \phi\right)^2 + \ldots 
$$
eliminates  all $x$-derivatives of $\phi$ and $W_\alpha$ (denoted by
$\ldots$) in terms of those of $\psi_\alpha$ using the nilpotency properties of
$\psi_\alpha$, $\bar\psi_{\dot\alpha}$  and the reduction formulas like 
$$
\bar \psi^2 (\partial \phi\cdot \partial \phi) = -{1\over 2}\psi^4 (\partial
\psi^\alpha\cdot \partial \psi_\alpha) \left(1 +{1\over 2}\bar D^2 \bar
\phi\right)^2~, \quad  \psi^4 \Box \phi = -\psi^4\,(\partial \psi^\alpha\cdot
\partial \psi_\alpha) \left(1 +{1\over 2}\bar D^2 \bar \phi\right)~, 
$$
which also follow from \p{phiExpr}, \p{Wexpr}. The same procedure is
applied to similar terms in \p{phiExpr}. As the next step, one substitutes 
$\psi^2 = W^2 \left( 1 +{1\over 2}\bar D^2 \bar \phi\right)^{-2} + \ldots$ 
into \p{phiExpr}. All ``superfluous'' terms  are canceled among themselves 
and one ends up with the simple  relations 
\be
\phi = -{1\over 2} \; \frac{W^2}{1 + {1\over 2} \bar D^2 \bar\phi}~, \quad 
\bar\phi = -{1\over 2} \; \frac{\bar W^2}{1 + {1\over 2} D^2 \phi}~,
\label{basicRel} \ee
which are just those postulated in \cite{BG} and derived from 
the nilpotency condition in \cite{RT}. We see that the same relations follow
from our generic procedure applied to the given specific case. An advantage
of this derivation is that it sets the direct relationship with the
``canonical'' nonlinear realization through the equations \p{phiExpr},
\p{Wexpr}. In particular, notice the relation 
$$
\psi^4 = \psi^2\bar\psi^2 = W^2\bar W^2 \left[\left(1 +{1\over 2}\bar D^2 \bar
\phi\right)\left(1 + {1\over 2} D^2 \phi\right)\right]^{-2}~.   $$

As was shown in \cite{BG,RT} the chiral superfield $\phi$ is just
the  Goldstone superfield Lagrangian density for the $N=2 \rightarrow N=1$
PBGS (it is the Fayet-Iliopoulos term from the $N=2$ perspective). It
describes a $N=1$ superextension  of the $d=4$ BI theory with  the second
hidden $N=1$ supersymmetry, or, equivalently, the gauge-fixed space-filling
D3-brane in a flat  background. For completeness, we quote here the solution 
of  \p{basicRel} \cite{BG}
\bea
&& \phi = -{1\over 2} \left\{W^2 + {1\over 2}\bar D^2 \,\frac{W^2\bar W^2}{1
-{1\over 2}A + \sqrt{1 - A +{1\over 4} B^2}} \right\} \label{N2BI} \\   
&& A \equiv {1\over 2}\left(D^2W^2 +\bar D^2\bar W^2\right)~, \quad  B \equiv
{1\over 2}\left(D^2W^2 -\bar D^2\bar W^2\right)~. 
\eea
Having at our disposal the explicit relations \p{phiExpr}, \p{Wexpr} we 
can in principle explicitly check, along the lines of Subsect. 2.2,  the
equivalence between the equations of motion corresponding to the $N=2
\rightarrow N=1$ BI Lagrangian \p{N2BI} and eqs. \p{chir5e3d} proposed
within the original nonlinear realization setting.   
\setcounter{equation}{0}

\section{Concluding remarks}
In this contribution we reviewed basic features of the PBGS approach to
superbranes and presented a few novel developments. In
particular, we showed that the method of constructing Goldstone superfield
actions which is based on the general relationship between linear and nonlinear
realizations of PBGS \cite{IvanovKapustnikov,DIK,IKLZ} is fairly workable not
only in the simple examples treated in this way earlier \cite{DIK,IKLZ}, but
also in some more complicated and interesting cases including the space-filling
D3-brane (the $N=2 \rightarrow N=1$ BI theory). In this short review we left
aside such interesting cases as the $N=4 \rightarrow N=2$ and $N=8 \rightarrow
N=4$  BI theories (super D3- and D6-branes in $D=6$ and $D=10$)
\cite{BIK4,BIK5} which  certainly offer new domains for applying the
machinery expounded here. A further work is also required in order to
understand in full the links between the PBGS and superembedding 
\cite{Sorokin} approaches. In recent papers \cite{PST,DH,BPPST}, the $N=1,\,
D=4$ supermembrane and D2-brane PBGS actions \p{lagrMem}, \p{membr1} and
\p{D2dens}, \p{muaction} originally derived in \cite{IK1} were recovered from
the superembedding approach, and some steps toward a similar derivation of
the PBGS D3-brane action of ref.\cite{BG} were undertaken. It would be tempting
to understand linear realizations of the PBGS theories and their relationship
to nonlinear realizations from the superembedding point of view.              

\section*{Acknowledgments}
I thank S. Bellucci and S. Krivonos in collaboration with whom some of
the results mentioned here were obtained. Useful discussions with them, as well
as with  F. Delduc,  A. Kapustnikov, O. Lechtenfeld, D. Sorokin, K. Stelle and
B. Zupnik are kindly acknowledged. I am grateful to the Organizers of the Misha
Saveliev Seminar in  Protvino for inviting me to give a talk and for kind
hospitality.  This work was supported in part by the RFBR-CNRS Grant No.
98-02-22034, RFBR Grant No. 99-02-18417 and NATO Grant No. PST.CLG 974874.

\end{document}